# Reply to "No spin-statistics connection in nonrelativistic quantum mechanics"

Murray Peshkin*

*Physics Division, Argonne National Laboratory, Argonne, Illinois 60439*

In a previous paper I gave an elementary proof, starting from stated assumptions of nonrelativistic quantum mechanics, that identical spin-zero particles must be bosons. Since then it has been suggested that my proof assumed its conclusion, and that it is based on a theory "quite different from standard physics." I show here that those two statements are incorrect.

My paper [1] reported a step toward the completion of a program initiated by Leinaas and Myrheim [2] and advanced by Berry and Robbins [3,4]. Suppose one implements the principle that the variables in nonrelativistic quantum mechanics should stand in one-to-one relation to the possible results of measurements by assuming that the argument of the wave function for two identical spinless particles must be a function of the unordered coordinate pair $\{\mathbf{r}_1, \mathbf{r}_2\}$, for which $\{\mathbf{r}_2, \mathbf{r}_1\}$ is the same point in the configuration space as is $\{\mathbf{r}_1, \mathbf{r}_2\}$, and similarly for identical particles with spin. Can one then deduce the connection between spin and statistics without recourse to relativistic quantum field theory? In different words, can one justify the usual procedure in which one assigns unobservable identity labels to the individual particles and then imposes the constraint that the wave function should be symmetric or antisymmetric under exchange of the two particles, according to their spin? Refs.[2-4] established important connections between spin and the behavior of wave functions under parallel transport but did not reach the spin-statistics connection. In Ref.[1], starting from the Leinaas-Myrheim assumption about the configuration space, I showed by elementary methods that the relative orbital angular momentum of pairs of identical spin-zero particles must have even integer values. Then the particles are bosons, not fermions. I was able to do that because I made use of an assumption not used by Refs.[2-4], that the wave functions must be continuous functions of $\mathbf{r}_1$ and $\mathbf{r}_2$ because of the second derivatives in the Hamiltonian. That and the Leinaas-Myrheim assumption that the configuration space is that of the unordered pairs $\{\mathbf{r}_1, \mathbf{r}_2\}$ are the only assumptions used in Ref.[1] that are not common to all accepted treatments of standard nonrelativistic quantum mechanics.

Allen and Mondragon, in "No spin-statistics connection in nonrelativistic quantum mechanics" [5], do not challenge the logic of Ref.[1] but they present two objections to its starting point: 1) The result is trivial. I started by assuming a wave function symmetric in $\mathbf{r}_1$ and $\mathbf{r}_2$, thereby assuming the stated conclusion that the spinless particles must be bosons, and 2) "The resulting theory is quite different from standard quantum mechanics." I will show here that those two assertions are false.

Two non-identical particles have the usual center-of-mass and relative coordinates $\mathbf{R} = (\mathbf{r}_1 + \mathbf{r}_2)/2$ and $\mathbf{r} = (\mathbf{r}_1 - \mathbf{r}_2)$. Their configuration space is the entire six-dimensional Euclidean continuum. For identical particles, however, $\mathbf{r}$ has to be identified with $-\mathbf{r}$. In other words, the relative coordinate has a magnitude and an axis, but not a direction along that axis. In Ref.[1], following the idea of Leinaas and Myrheim, I avoided the complicated multiply-connected topology for the identical particles by defining the relative coordinate as

$$\mathbf{r} = \mathbf{r}_2 - \mathbf{r}_1 \quad \text{where} \quad \begin{cases} z_2 > z_1, \text{ or} \\ z_2 = z_1 \text{ and } y_2 > y_1, \text{ or} \\ z_2 = z_1 \text{ and } y_2 = y_1 \text{ and } x_2 > x_1. \end{cases} \tag{1}$$

$\mathbf{r} = \mathbf{r}_1 - \mathbf{r}_2$ elsewhere.

The domain of the so-defined $\mathbf{r} = (x, y, z)$ is a simply-connected space including positive $z$ and half of the $z=0$ plane. The points $(\mathbf{R},\mathbf{r})$ are in one-to-one correspondence with the unordered pairs $\{\mathbf{r}_1, \mathbf{r}_2\}$ and a vector and its negative are never both contained in the domain of $\mathbf{r}$. This domain of $\mathbf{r}$ differs from the ones defined in Refs.[2-4] in that here the one-to-one correspondence between $\{\mathbf{r}_1, \mathbf{r}_2\}$ and $(\mathbf{R},\mathbf{r})$ is achieved without identifying certain pairs of points in such a way that the domain of $\mathbf{r}$ becomes multiply connected, but at the price of creating discontinuities in $\{\mathbf{r}_1, \mathbf{r}_2\}$ as function of $(\mathbf{R},\mathbf{r})$ in the $z_1 = z_2$ plane. (Take $\mathbf{r}_2 = -\mathbf{r}_1 = (a, 0, \varepsilon)$. For $\varepsilon > 0$, $x = 2a$, but for $\varepsilon < 0$, $x = -2a$.) Those discontinuities are prevented from violating the physics by the assumption that wave functions must nevertheless be continuous functions of $\mathbf{r}_1$ and $\mathbf{r}_2$.

In the domain of $\mathbf{r}$, wave functions can be written as

$$\psi(\mathbf{R},\mathbf{r}) = \sum_{\ell m} a_{\ell m}(\mathbf{R},r) Y_{\ell m}(\hat{\mathbf{r}}). \tag{2}$$

Here, in contrast to the situation in the full three-dimensional space, the sets $Y_{\ell m}(\hat{\mathbf{r}})$ for even and for odd $\ell$ are separately complete. It was shown in Ref.[1] that even values of the relative orbital angular momentum $\ell$ are superselected from odd. Both cannot appear in a wave function.

The Comment observes correctly that the wave function in Eq.(2) is symmetric under $\mathbf{r}_1 \leftrightarrow \mathbf{r}_2$ and concludes that the identical particles must have been made bosons by assumption. The symmetry under $\mathbf{r}_1 \leftrightarrow \mathbf{r}_2$ is true, but the conclusion is not true in the present case, where the domain of $\mathbf{r}$ is that given in Eq.(1).

In the usual treatments, where the two spin-zero particles are distinguishable, or where they are treated as distinguishable by labeling them in defiance of the principle that the configuration variables should represent observables, the domain or $\mathbf{r}$ is the entire three-dimensional space. Terms with even $\ell$ are symmetric under exchange of the two particles, *i.e.* under $\mathbf{r}_1 \leftrightarrow \mathbf{r}_2$ or equivalently under $\mathbf{r} \to -\mathbf{r}$. Terms with odd $\ell$ are antisymmetric. If one goes on to insist, somewhat paradoxically, that all observables must be symmetric under $\mathbf{r} \to -\mathbf{r}$, even $\ell$ are superselected from odd because no observable connects them. The particles are bosons with even $\ell$ or fermions with odd $\ell$, but the choice between those two possibilities has to be imposed by assumption or by making use of relativistic quantum field theory. In that situation $\psi(\mathbf{R},\mathbf{r}) = (-1)^\ell \psi(\mathbf{R},-\mathbf{r})$. The wave function is even or odd under exchange of the two particles according to whether one allows only even or only odd $\ell$.

Here, by contrast, the configuration space for identical spin-zero particles is taken to be the unordered pairs $\{\mathbf{r}_1, \mathbf{r}_2\}$ with $\mathbf{r}$ defined by Eq.(1). Then the wave function is always symmetric under $\mathbf{r}_1 \leftrightarrow \mathbf{r}_2$ but that does not imply even $\ell$. The domain of the relative $\mathbf{r}$ never includes $-\mathbf{r}$. The wave function $\psi(\mathbf{R},-\mathbf{r})$ does not exist and the reasoning applied above to distinguishable particles cannot be applied to identical particles. Both odd and even values of $\ell$ are possible in spite of the symmetry under $\mathbf{r}_1 \leftrightarrow \mathbf{r}_2$. For example, the wave function

$$\psi(\mathbf{r}) = (x + iy) e^{-r^2}, \tag{3}$$

has $\ell = 1$, but it is nevertheless symmetric under $\mathbf{r}_1 \leftrightarrow \mathbf{r}_2$ for identical particles, as are all functions of the unordered $\{\mathbf{r}_1, \mathbf{r}_2\}$. Therefore, symmetry of the wave function does not imply boson statistics for identical particles.

It was shown in Ref.[1] that an additional assumption of continuity eliminates the odd values of $\ell$, proving that identical spinless particles can only be bosons, not fermions. The example of Eq.(3) illustrates that general result. It is not an acceptable wave function under the assumptions of Ref.[1] because it has discontinuities as a function of $\mathbf{r}_1$ and $\mathbf{r}_2$ in the $z_1 = z_2$ plane.

The Comment additionally states that Ref.[1] results in a theory different from standard quantum mechanics. In fact, the theory is unchanged. In standard quantum mechanics with boson statistics, all wave functions and all operators that represent observables are unchanged when the relative $\mathbf{r} \rightarrow -\mathbf{r}$. Then all matrix elements of the observables are unchanged if one restricts the integrals to values of $\mathbf{r}$ having positive $z$, appropriately correcting the normalization of course. That procedure is gives precisely the matrix elements that appear when the domain of $\mathbf{r}$ is defined as in Ref.[1]. Standard quantum mechanics is justified, not replaced, by Ref.[1].

If one chooses to allow only odd $\ell$, in violation of the present assumptions for spin-zero particles because of the discontinuous wave functions, the same argument shows that one obtains standard quantum mechanics with fermion statistics.

Finally, I note that the Comment, in and around its Eq.(7), states that the assumptions of Ref.[3] result in the connection between spin and statistics for all values of the spin. Berry and Robbins did say that in Ref.[3], but they later showed in Ref.[4] that the proof was not valid and that particles of any spin can be bosons or fermions under their assumptions.

I thank Alexander Volya for critical reading of this manuscript and Roland Allen for stimulating correspondence. This work was supported by the U.S. Department of Energy, Nuclear Physics Division, under Contract No. W-31-109-ENG-38.

---

*Electronic address: peshkin@anl.gov